\documentclass[aps,prb,twocolumn,groupedaddress,showpacs,floatfix,altaffilletter]{revtex4-1}
\usepackage{graphicx}
\usepackage{grffile}
\usepackage{amsmath}
\usepackage{amssymb}
\usepackage{bm}
\usepackage{color}
\usepackage[dvipsnames]{xcolor}
\usepackage{amsmath,amssymb,amsfonts}
\usepackage{epsfig}
\usepackage{times}
\usepackage[colorlinks,bookmarks=false,citecolor=blue,linkcolor=red,urlcolor=blue]{hyperref}

\newcommand{\fref}[1]{Fig.~\ref{#1}}
\newcommand{\eref}[1]{Eq.~(\ref{#1})}

\newcommand{\smallwidth}{0.6\columnwidth}
\newcommand{\figwidth}{0.97\columnwidth}

\newcommand{\midwidth}{0.75\columnwidth}

\begin{document}

\title{Temperature dependence of NMR Knight shift in pnictides: proximity to a van Hove singularity }

\author{R.~Nourafkan, S. Acheche}
\affiliation{Institut quantique \& D{\'e}partement de Physique and RQMP, Universit{\'e} de Sherbrooke, Sherbrooke, Qu{\'e}bec, Canada  J1K 2R1}
\date{\today}
\begin{abstract}
The unconventional temperature variation of the Knight shift (static spin susceptibility)  that has been observed  in Fe-based superconductors AFe$_2$As$_2$ (A = K, Rb, Cs) is explained in terms of proximity to a van Hove singularity. Using the Hubbard model we show that when the Fermi energy is in the vicinity of a van Hove singularity, a downturn in spin susceptibility occurs as the temperature is lowered. This behavior is  characterized by a temperature, $T^*$, which is determined by the difference in energy between the Fermi level and the van Hove singularity. When vertex corrections are taken into account in a dynamical mean-field approximation, the effect of correlations amplifies the relative drop in the Knight shift and moves $T^*$ to lower temperatures.
\end{abstract}

\pacs{74.70.Xa, 71.27.+a}

\maketitle
Nuclear magnetic resonance (NMR) techniques provide a probe of the spin response at specific atomic locations.
In an itinerant electron system, the spin part of the Knight shift measured in NMR experiments is proportional to the uniform spin susceptibility, $K_S(T) = B \chi^m(T)$ where $B$ denotes the hyperfine coupling describing coupling between nuclear spins and itinerant electron spins. The hyperfine coupling is temperature independent, hence, the temperature dependence of the Knight shift, $K_S$, is identical with that of  spin susceptibility. The spin susceptibility of  itinerant electrons is given by the Pauli susceptibility. For non-interacting systems, $\chi^m$ takes the form $(1/4)\int d\epsilon \rho(\epsilon) (dn(\epsilon)/d\epsilon)$ where $n(\epsilon)$ is the Fermi distribution function and $\rho(\epsilon)$ denotes the total density of states. It depends weakly on the temperature and upon decreasing temperature smoothly saturates to its $T=0$ limit, \emph{i.e.}, $\sim \rho(\epsilon_F)/4$, where $\epsilon_F$ is the Fermi energy. 

In heavy Fermion systems with both localized $f$ electrons and itinerant conduction electrons $c$, the temperature dependence of the Knight shift may differ from the temperature dependence of the total spin magnetization. This so-called Knight shift anomaly can be understood in terms of two hyperfine couplings to the two different electron spins (localized vs itinerant)~\cite{ShirerE3067}. Then the Knight shift is given by $K_S=B_1\chi_{cc}^m+(B_1+B_2)\chi_{cf}^m+B_2\chi_{ff}^m$, \emph{i.e.} Knight shift weighs the different correlation functions separately.  At temperatures higher than a material-dependent charectristic temperature, $T >T^X$, the Curie-Weiss susceptibility of the local moments dominates the temperature-independent Pauli susceptibility of the conduction electrons, then $K_S \simeq B_2\chi_{ff}^m$. Therefore,  $K_S$  monotonically increases upon decreasing $T$ for these values of temperature. Below $T^X$,  $\chi_{cf}^m$ becomes significant and governs the temperature dependence of the Knight shift, which is different from the Curie-Weiss law~\cite{ShirerE3067}.

Recently, a similar Knight shift anomaly (crossover) was observed in heavily hole-doped Fe-based superconductors AFe$_2$As$_2$ (A = K, Rb, Cs); at low temperature the Knight shift deviates from a Curie-Weiss behavior describing the high temperature regime~\cite{PhysRevLett.116.147001,  PhysRevLett.111.027002}. A similar behavior is seen for the spin susceptibility of KFe$_2$As$_2$~\cite{PhysRevLett.111.027002}. The characteristic crossover temperature, $T^*$, decreases continuously when K is substituted with the larger alkaline ions Rb or Cs.  Below $T^*$, the Knight shift decreases upon decreasing $T$ and eventually saturates at very low temperature. Due to the similarity and observed large effective masses in these compounds, it was suggested that the Knight shift crossover in AFe$_2$As$_2$ can indicate an orbital-selective Mott transition in which electrons in the $d_{xy}$ orbital undergo a Mott transition and become localized  while electrons in $d_{xz}$ and $d_{yz}$ remain itinerant~\cite{PhysRevLett.116.147001, PhysRevLett.111.027002, PhysRevB.97.045118, PhysRevLett.112.177001}.  However, this scenario is highly debated for iron-based superconductors which are believed to be Hund's metal with the multiorbital nature as the key factor~\cite{1367-2630-11-2-025021,  doi:10.1146/annurev-conmatphys-020911-125045, Yin2011}.

Here we propose an alternative explanation of this behavior.  Indeed, a van Hove singularity (vHS) has been observed in angle-resolved photoemission spectroscopy (ARPES) of  AFe$_2$As$_2$ and confirmed by LDA calculation~\cite{PhysRevB.92.144513}. The vHS is located just a few meV below the Fermi level and moves towards it upon substitution of K with  Rb or Cs. The proximity of the vHS proximity can induce a pronounced  temperature dependence of the Pauli susceptibility. It has also been proposed as responsible for both the heavy mass behavior observed in these materials, and for their superconducting gap symmetry~\cite{PhysRevB.92.144513, 1807.00193}. Here, we show that the Knight shift shows a similar crossover due to the proximity of the vHS. We show that the characteristic temperature $T^*$ scales with the difference in energy between the Fermi level and the position in energy of the vHS, $\epsilon_{vHS}$; it moves to higher temperature upon increasing this energy difference. Furthermore, $K_S(T^*)-K_S(T\rightarrow 0)$ decreases when the Fermi level is located further away from $\epsilon_{vHS}$. We also investigate the effect of electron-electron interactions on this behavior. We find that upon increasing electron-electron interaction, $K_S(T^*)-K_S(T\rightarrow 0)$ increases and $T^*$ shifts to lower temperatures. 

\begin{figure}
 \includegraphics[width=\midwidth]{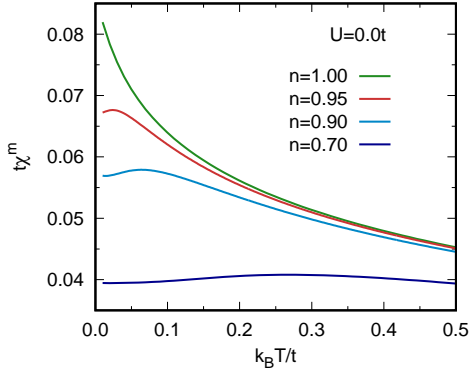} 
\caption{ Knight shift $K_S \propto \chi^{m}$ of the non-interacting system as a function of temperature $k_BT/t$ for several electron densities. The peak position of the Knight shift moves to higher temperature and become less pronounced  when Fermi energy moves away from vHS energy location.}
 \label{fig:KnightShiftNonInt}
\end{figure}

{\it Model and method --} The influence of a vHS on the Knight shift can be discussed using the Hubbard model on the square lattice, 
\begin{equation}
H=-t\sum_{\langle ij \rangle \sigma}  c^{\dagger}_{i\sigma}c_{j\sigma}+U\sum_i n_{i\uparrow}n_{j\downarrow},\label{eq:Hubbard}
\end{equation} 
where $c^{\dagger}_{i\sigma}$($c_{i\sigma}$) creates (annihilates) an electron with spin $\sigma$ on site $i$ and $n_{i\sigma}=c^{\dagger}_{i\sigma}c_{i\sigma}$. The amplitude $t$ denotes the nearest-neighbor hopping amplitude,  and $U$ the on-site screened Coulomb interaction. The non-interacting density of states of this model posseses a vHS at zero energy. At half-filling the Fermi energy lies on the $\epsilon_{vHS}$.
This model also allows us to discuss the impact of correlations on the temperature dependence of the Knight shift. The model has particle-hole symmetry, hence similar results would be obtained whether the vHS is located above or below Fermi level.

In general, the Pauli susceptibility is determined by the ${\bf q}\rightarrow 0$ and $\nu_n\rightarrow 0$ limit of ${\chi}_{ph}^{m}({\bf q},\nu_n)$ 
where ${\chi}_{ph}^{m}$ is the lattice magnetic susceptibility.  In an interacting system, the so-called generalized dressed spin susceptibility can be calculated from the Bethe-Salpeter equation  as~\cite{Bickers2004, PhysRevLett.117.137001}
\begin{align}
{\bm \chi}^{m}(Q) &=\left[ {\bm 1} - { \bm\chi}^{0}_{ph}(Q)
{ \bm\Gamma}^{m,irr}(Q)\right]^{-1}{\bm\chi}^{0}_{ph}(Q).
\label{eq:BSSus}
\end{align}
where bold quantities are matrices. The bubble susceptibility is defined as
\begin{equation}
[{\chi}^{0}_{ph}(Q)]_{K,K'} =-(N\beta) { G}(K+Q) { G}(K)\delta_{K,K'}.  \label{BubbleSus}
\end{equation}
Here, ${ G}(K)$ is the dressed particle propagator, $K\equiv ({\bf k},i\omega_m)$ denotes momentum/energy four-vectors (the lattice is two-dimensional), $N$ is number of ${\bf k}$-points and $\beta = 1/(k_BT)$.  In \eref{eq:BSSus}, ${\bm \Gamma}^{m,irr}$ is the irreducible vertex function describing the irreducible interaction of the two elementary excitations. \eref{eq:BSSus} is the common part of the response to an external field and solely depends on the electronic structure of the system. 
An observable response function, on the other hand, is obtained by closing the external legs of \eref{eq:BSSus} using appropriate oscillator matrix elements, $O(Q)$ and  $O(-Q)$, 
 \emph{i.e.}, 
\begin{equation}
\chi_{\rm obs}^{m}(Q) = \frac{1}{N^2\beta^2} \sum_{KK'} O_{K,Q} [{ \chi}^{m}(Q)]_{KK'}O_{K',-Q}. \label{Obs}
\end{equation}
The oscillator matrix elements depend on the orbital wave-function and the field wave-vector and frequency.\cite{PhysRevB.96.125140}  In the ${\bf q}\rightarrow {\bm 0}$ and $\nu_n\rightarrow 0$ limit, with an orthonormal basis set, the oscillator matrix element in the magnetic channel of a single-band system reduces to the identity multiplied by $1/2$ due to the definition of spin in terms of electron densities, \emph{ i.e.}, $S^z=(n_{\uparrow}-n_{\downarrow})/2$.

We solve the Hamiltonian, \eref{eq:Hubbard},  using DMFT
and the exact diagonalization (ED) method~\cite{PhysRevLett.72.1545}.  In general, $[{\bm \Gamma}^{m,irr}(Q)]_{K,K'}$ depends on the the transferred momentum/frequency in a scattering process, $Q$, and on the incoming momentum/frequency variables. The out-coming variables are determined by conservation laws.  In a normal system, there
is a range and a characteristic relaxation time, beyond which
$[{\bm \Gamma}^{m,irr}(Q)]_{K,K'}$ becomes negligible, Hence, the spatially local part of
the irreducible vertex function is the dominant part. This part of the irreducible vertex function, $\big[{ {\bm \Gamma}}^{m, irr}_{loc}(\nu_n)\big]_{\omega_m\omega_{m'}}$, can be calculated in the framework of the DMFT approximation from four point correlation functions on the self-consistent impurity~\cite{PhysRevB.75.045118,PhysRevB.86.125114, 1807.03855}.  
A common approximation consists in substituting  the irreducible vertex function by ${\bm \Gamma}^{m, irr}_{loc}(\nu_n)$ and neglecting the non-local part~\cite{RevModPhys.68.13}. The DMFT(ED) algorithm is also used to compute the local part of the irreducible vertex function~\cite{RevModPhys.68.13,PhysRevB.86.125114, PhysRevB.75.045118}.   

\begin{figure}
 \includegraphics[width=\figwidth]{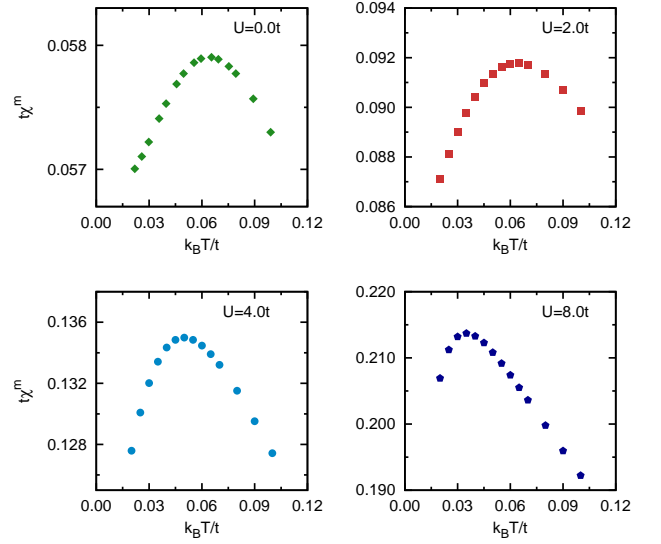} 
\caption{ Knight shift $K_S \propto \chi^{m}$ as a function of temperature $k_BT/t$ for $U=0.0 t$, $U=2.0 t$, $U=4.0T$ and $U=8.0 t$.  The electron density is $n=0.9$. The peak position of the Knight shift moves to lower temperature upon increasing $U$.}
 \label{fig:KnightShift}
\end{figure}

{\it Results -- } Figure~\ref{fig:KnightShiftNonInt} shows the Knight shift $K_S \propto \chi^{m}$ of the non-interacting system as a function of temperature for several electron densities.  At $T \rightarrow 0$, the Knight shift saturates to $\rho(\epsilon_F)/4$. Upon increasing temperature, the thermal function $(dn(\epsilon)/d\epsilon)$ broadens, leading to a finite contribution of the vHS to the spin susceptibility. Hence, intially, the spin susceptibility upon increasing temperature, exhibits a broad maximum at $T^*$ and then monotonically decrease beyond, following approximately a Curie-Weiss law for higher temperatures. The high-$T$ reduction in the magnetic susceptibility is due to fast dynamics of electron spins. 
The maximum in the Knight shift becomes more pronounced  and  occurs at lower temperature for larger electron densities, namely, when the Fermi level, $\epsilon_F$, approaches the vHS $\epsilon_{vHS}$. At $n=1$, where the Fermi energy lies on the vHS energy, the maximum occurs at $T=0$. Even in the non-interacting level, this trend is consistent with experimental results on AFe$_2$As$_2$ where $T^*$ is the smallest for the Cs compound with the smallest energy difference $\epsilon_{F}-\epsilon_{vHS}$. 

An interacting system is more polarizable than a non-interacting one. A Fermi-liquid system, for instance, exhibits an enhanced Pauli susceptibility given by $(1+F^a_0)^{-1}\chi_0$ where  $F^a_0 <0$ is Landau parameter.  On the other hand, interactions broaden the vHS. Moreover, the response of an interacting system is not restricted to the electrons at the Fermi level but electrons around it also contribute. This raises the question of the impact of interactions on the above picture. Here, we restrict ourselves to  weak to intermediate interaction strengths, which is appropriate for iron-based superconductors. The temperature dependence of $K_S$ for $U=2.0t$, $U=4.0t$, and $U=8.0t$ are shown in \fref{fig:KnightShift} and compared with the non-interacting case. As expected, the spin susceptibility is enhanced by interactions. Moreover, the downturn of $\chi^m$ at low $T$ becomes more pronounced.  The characteristic temperature, $T^*$, moves to lower temperatures upon increasing $U$. It is likely that the saturation of the spin susceptibility at very low $T$ occurs at lower temperatures as interaction strength is increased. It is expected that AFe$_2$As$_2$ (A = K, Rb, Cs) compounds have similar interaction strengths, therefore the characteristic temperature is mainly determined by $\epsilon_F-\epsilon_{vHS}$. 

\begin{figure}
 \includegraphics[width=\figwidth]{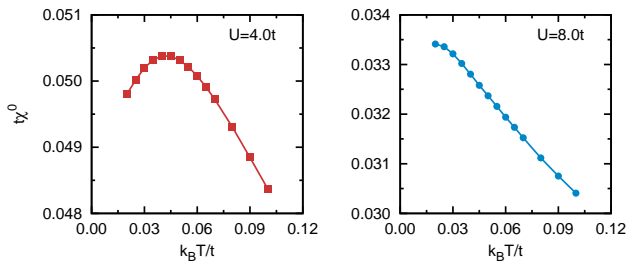} 
\caption{ Dressed bubble susceptibility as a function of temperature $k_BT/t$ for $U=4.0t$ (left) and $U=8.0 t$ (right).  The electron density is $n=0.9$. At larger $U$ values, the dressed bubble susceptibility does not show a downturn near the $T^*$ calculated including vertex corrections. }
 \label{fig:Bubble}
\end{figure}

Since evaluation of the irreducible vertex function is difficult, in real material calculations the spin susceptibility is often approximated with the dressed bubble diagram, \eref{BubbleSus}. However, our calculations show that at large interaction strengths, the temperature dependence of the bubble susceptibility is different from the susceptibility calculated with vertex corrections. As can be seen form \fref{fig:Bubble}, in contrast to $U=4.0t$ where the downturn of the spin susceptibility is present at the bubble level, for $U=8.0t$ the bubble susceptibility increases upon decreasing $T$ and does not show a downturn near the $T^*$ calculated in \fref{fig:KnightShift}, which includes vertex corrections. Therefore, it is essential to take into account vertex corrections for large values of the interaction to obtain the correct temperature dependence.

Furthermore, it is also customary to inspect the temperature dependence of the impurity susceptibility instead of $\chi^m({\bf q}={\bm 0}, \nu_n=0)$. Our results show that the downturn of $\chi^m({\bf q}={\bm 0}, \nu_n=0)$ cannot be seen from the impurity susceptibility. This can be understood if one assumes that, upon decreasing temperature, the dressed susceptibility at non-zero momenta grows faster than the reduction of $\chi^m({\bf q}={\bm 0})$, hence, the local susceptibility, obtained from summation over all momenta, does not show the downturn seen in $\chi^m({\bf q}={\bm 0})$. 
\begin{figure}
 \includegraphics[width=\smallwidth]{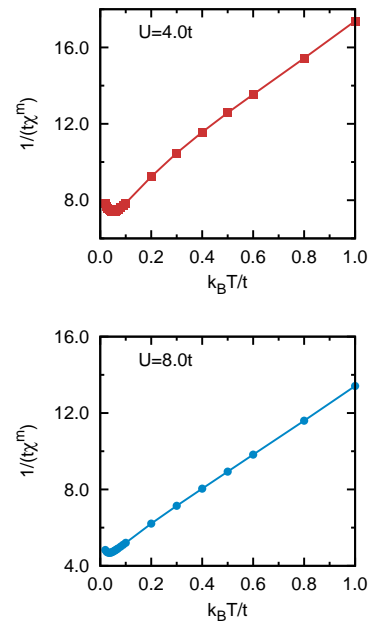} 
\caption{ Inverse spin susceptibility as a function of temperature $k_BT/t$ for $U=4.0t$ (top) and $U=8.0 t$ (bottom).  A Curie-Weiss law, $\chi_m^{-1} (T) \propto T+\theta$, becomes clearly established even at relatively low temperature (on electronic scales) for $U=8.0t$. The electron density is $n=0.9$. }
 \label{fig:Curie-Weiss}
\end{figure}

Figure~\ref{fig:Curie-Weiss} displays the inverse spin susceptibility as a function of temperature. As can be seen from the figure, the temperature dependence of the spin susceptibility (Knight shift) at high temperature is consistent with a Curie-Weiss behavior, $\chi^m (T) \propto (T+\theta)^{-1}$.  At very high temperature, of order of the bandwidth, the spin susceptibility approaches its value for localized non-interacting spins,  \emph{i.e.}, $1/(4T)$ (not shown).  A Curie-Weiss law, suggesting a local-moments dominated behavior, holds down to a lower temperature upon increasing interaction strength. As the temperature is decreased the susceptibility crosses-over from Curie-Weiss behavior to Fermi liquid behavior with a pronounced temperature dependence due to the proximity of the vHS. 

The spin-lattice relaxation rate $1/(T_1T)$  probes the low-frequency behavior of the spin susceptibility on the real axis. In a Fermi liquid state, a Korringa-like relaxation $1/(T_1 T) \sim $~const. is expected, whereas in a localized spin system $1/(T_1 T) \sim (T+\theta)^{-1}$. The experimental spin-lattice relaxation rates for AFe$_2$As$_2$ show a power-law dependence on temperature,  $1/(T_1T) \propto T^{-\eta}$. However, the exponent changes around $T^*$:  for $T<T^*$,  $\eta \simeq 0.25$ while $\eta \simeq 1$ for $T>T^*$, although there are not enough data points for  $T>T^*$ to be conclusive~\cite{PhysRevLett.116.147001}.  

 When the wave vector-dependence of the hyperfine interaction is neglected, the spin relaxation rate is given by~\cite{doi:10.1143/PTP.16.23}
\begin{align}
\frac{1}{T_1T} \propto \lim\limits_{\nu \rightarrow 0} (\frac{1}{N}) \sum_{\bf q}  \frac{\textrm{Im} \chi({\bf q}, \nu)}{\nu}.
\label{relaxationTime}
\end{align}
%
We use Pad\'e analytic continuation for the impurtity susceptibility at $U=4.0t$, which is the best DMFT approximation for the local susceptibility. We find that $1/T_1T$ is almost temperature-independent for $T<aT^*$ while it decreases upon increasing temperature for $T>aT^*$ (not shown), where $a$ is a multiplicative factor slightly larger than unity. We believe $a \neq 1$ may be an artefact of the analytic continuation. Indeed, we can also analytically continue to zero-frequency using the approximation $(1/N) \sum_{\bf q}  \chi({\bf q}, \tau=1/2T)/(\pi T^2)$, where $\tau$ denotes imaginary time~\cite{PhysRevLett.69.2001}. This is correct if $\textrm{Im} \chi({\bf q}, \nu)/\nu$ remains frequency independent for $\nu < 2T$. This condition is not fully satisfied here. However, by employing this equation we find a change in $1/T_1T$ temperature dependence behavior at $T^*$. Therefore, the relaxation rate temperature-dependence changes around the characteristic temperature in agreement with experimental results, however, $\eta$ values do not fully agree. In our calculation,  $1/T_1T \sim 1/(T+\theta)$ for $T>T^*$.

{\it Conclusion --} Using the Hubbard model on the two-dimensional square lattice, we showed that  a downturn in temperature-dependence  of the spin susceptibility takes place with a characteristic temperature $T^*$.  The characteristic temperature scales with the difference in energy between the Fermi level and the van Hove singularity.  When vertex corrections are included with the DMFT-dressed propagators, the effect of the van Hove singularity seen in the non-interacting case is amplified and $T^*$ moves to lower temperatures. Hence, given ARPES data on the proximity between the van Hove singularity and the Fermi level in AFe$_2$As$_2$ (A = K, Rb, Cs), this could naturally explain  the  main qualitative  features  of  the  measured Knight  shift,  without  appeal  to  an orbital-selective Mott transition. 

\begin{acknowledgments}
We are deeply indebted to A.-M.S. Tremblay for insightful discussions and for careful reading of the manuscript. R.~N is thankful to P.~Richard for useful discussions.   This work has been supported by the the Canada First Research Excellence Fund, the Natural Sciences and Engineering Research Council of Canada (NSERC) under grant RGPIN-2014-04584, and by the Research Chair in the Theory of Quantum Materials. Simulations were performed on computers provided by the Canadian Foundation for Innovation, the Minist\`ere de l'\'Education des Loisirs et du Sport (Qu\'ebec), Calcul Qu\'ebec, and Compute Canada.
\end{acknowledgments}

%

\end{document}